\documentclass[12pt,psfig]{article}
\usepackage{amssymb}
\usepackage{amsmath,amsthm}
\usepackage{amsfonts}
\usepackage{graphicx}
\usepackage{graphics}
\usepackage{indentfirst}
\usepackage{color,graphicx}
\usepackage{caption}
\usepackage[utf8]{inputenc} 
\usepackage[english]{babel}
\usepackage[nottoc]{tocbibind}
\usepackage[usenames,dvipsnames]{xcolor}
\usepackage[symbol]{footmisc}
\numberwithin{equation}{section}

\usepackage{hyperref}

\begin{document}
\begin{center}\Large\textbf{Observational Restrictions and
Slow-Roll D-brane Inflation in the Special $F(\phi,T)$ Gravity}
\end{center}
\vspace{0.75cm}
\begin{center}{\large Feyzollah Younesizadeh and \large Davoud
Kamani} {\footnote{\textcolor{Magenta}
{Corresponding author}}}
\end{center}
\begin{center}
\textsl{\small{Department of Physics, Amirkabir University of
Technology (Tehran Polytechnic) \\
P.O.Box: 15875-4413, Tehran, Iran \\
e-mails: fyounesizadeh@aut.ac.ir , kamani@aut.ac.ir \\}}
\end{center}
\vspace{0.5cm}

\begin{abstract}

We shall investigate the inflation for the D-brane model,
motivated by the modified gravity $F(\phi,T)$.
This gravity has been recently introduced in the literature.
The feasibility of the D-brane inflation theory
in the $F(\phi,T)$-gravity has been studied in 
conjunction with the most recent Planck
data. We shall analyze the slow-roll inflation 
in the context of the $F(\phi)T$-gravity,
via the D-brane model. Then, we shall 
calculate the inflation dynamics to
obtain the scalar spectral index ``$n_s$'' and the
tensor-to-scalar ratio ``$r$''. Besides, we 
investigate the dynamics of the reheating for 
this model. Our model accurately   
covers the left-hand side of the Planck data
and the D-brane inflation.

\end{abstract}


\newpage
\section{Introduction}

The spatial flatness and the scale-invariant 
density perturbations suggest
initial conditions during the exponential 
inflation, known as the visible inflation.
Research on the cosmic microwave background
(CMB) confirms the universe's flatness 
and the uniformity, which indicate an
accelerated expansion in the 
early universe \cite{1}-\cite{4}.
Combining the inflaton field $\phi$
with the Einstein-Hilbert 
action describes this inflationary period.
Thus, this preserves the original inhomogeneity
that leads to the observed large-scale structures.

The slow-roll inflation is an effective mechanism 
in which the potential of the inflation $V(\phi)$ drives 
the energy density during the 
inflation. The inflaton undergoes the slow-roll
with a nearly constant potential.
Achieving this flat potential strongly poses some 
challenges in the models with the  
particle physics basis \cite{5}, \cite{6}.

The recent data from Planck, BICEP and 
Keck Array \cite{7}-\cite{9} prefer  
concave inflationary potentials over convex
ones. The hilltop models of the inflation usually utilize 
concave potentials \cite{10}-\cite{17}. 
Squaring the hilltop potential is a simple 
solution in the symmetry-breaking of the 
inflation models \cite{11}, \cite{15}-\cite{18}.
Recent analysis favors models with the  
negative curvature potentials
over the hybrid and large field models \cite{19}. 
These models were found in the particle physics 
models with the symmetry breaking.

After the discovery of the D$p$-branes \cite{20},
the D-brane universe, with the string theory principles,
was introduced. String theory obviously identifies the dilaton 
field $\phi$ as the inflaton. 
The inflationary scale relative to 
the string theory scale can clarify the
lack of the fine-tuning in the e-folds of growth.
The inflation may originate 
from other dimensions, which was 
suggested by the D-brane inflation. 
The D-brane inflation models \cite{21}-\cite{25}
effectively utilize potentials that can be complements
to the $\alpha$-attractors \cite{26}-\cite{30}.
The analysis of the Planck's 2018 data prominently  
reveals the D-brane inflation models
with the string-theoretical and phenomenological elements
\cite{31}. 

The power-law potentials are popular inflationary
models with a single monomial potential \cite{32}, 
where the inflation
occurs at high values of the inflaton field.
Models that are inspired by the particle
physics, e.g., the natural inflation \cite{33}, \cite{34}
and the periodic potentials, contain
a pseudo-Nambu-Goldstone boson as the inflaton.
Other models that contain the Starobinsky $R^2$-inflation
\cite{35}, \cite{36} and the exponential
tails \cite{37} have been introduced via the 
supergravity and string theory. 

The main key in the inflationary process is the
development of the general relativity action through the 
geometric part or matter component. This has led to various
modified gravities with $f(R)$, $f(T)$, $f(G)$, $f(R, T)$,
and so on \cite{38} - \cite{48}. 
Among the appealing modified gravities, the 
$F(\phi)T$-gravity has the non-minimal
coupling of the scalar field with the trace of
the energy-momentum tensor ``$T$'' \cite{49}. 
This type of coupling term is generally motivated
by the quantum gravity. It provides
a comprehensive description of the unified gravity
with other fields \cite{50}-\cite{53}. 
One of the appeals of the 
$F(\phi)T$-gravity is that at the end of the inflation,
i.e., when the inflaton is decayed,
the Einstein's gravity is naturally returned. 
In fact, the $F(\phi) T$-gravity is
an extension of the $f(R,T)$-gravity.                
The latter one presents attractive results 
for the cosmic inflation \cite{54}, \cite{55}.
In Ref. \cite{48} the slow-roll inflation with
$F(\phi)T=\sqrt{\kappa}\;\phi T$ has been studied.
It revealed that the Natural, Chaotic,
and Starobinsky potentials behave in better agreement
with the data. 

All the foregoing facts, evidences and
affirmative conclusions stimulated and 
motivated us to adopt the $F(\phi)T$-gravity.
At first, we apply a general functional $F(\phi)$ 
and an arbitrary scalar potential $V(\phi)$. Afterward, 
special forms of the functionals $F(\phi)$ and $V(\phi)$ 
will be taken into account.
 
This paper is organized as follows. In Sec. \ref{300}, 
we shall compute inflationary effects arising from 
the $F(\phi)T$ term and a general potential $V(\phi)$, 
deriving the observables and slow-roll parameters. 
In Sec. \ref{400}, we explore the D-brane inflation 
by using specific forms of $F(\phi)$ and $V(\phi)$, 
for comparing our results with the Planck 2018 data. 
Additionally, we shall analyze the reheating dynamics and 
present relevant figures. 
Finally, Section \ref{500} is devoted to the conclusions 
and results.

\section{The inflation via the modified gravity $F(\phi)T$}
\label{300}

Here, we investigate the inflation via the 
modified gravity that emerges from
the interaction between a scalar field and the 
trace of the energy-momentum tensor.
Thus, we start with action \cite{48},
\begin{equation}
\label{16}
S=\int{{\rm d}^4x\sqrt{-g}}\left(\frac{R}{2\kappa}
+\beta F(\phi)g^{\mu\nu}T_{\mu\nu}+\mathcal{L}_{m}\right),
\end{equation}
where we apply 
$\kappa=8\pi G=1/M_{\rm P}^2$, in which $M_{\rm P}$ is
the Planck mass. The functional 
$\mathcal{L}_{m}$ represents the matter
Lagrangian, which is given as follows
\begin{equation}
\label{18}
\mathcal{L}_{m}=-\frac{1}{2}g^{\mu\nu}\partial_{\mu}
\phi\partial_{\nu}\phi-V(\phi).
\end{equation}
Besides, $T = g^{\mu\nu}T_{\mu\nu}$
is the trace of the energy-momentum tensor, 
which is extracted
from $\mathcal{L}_{m}$. For the canonical scalar field
it is given by
\begin{equation}
T=g^{\mu\nu}T_{\mu\nu}=-g^{\mu\nu}\frac{2}{\sqrt{-g}}
\frac{\partial(\sqrt{-g}\mathcal{L}_{m})}{\partial g^{\mu\nu}}
=-g^{\mu\nu} \partial_{\mu}\phi\partial_{\nu}\phi
-4 V(\phi).
\end{equation}

The functional $F(\phi)$ 
should obviously satisfy the condition $F(0)=0$ to
ensure that after decaying of the inflaton field 
$\phi$ we receive the Einstein's theory of gravity.
In this section, the functional $F(\phi)$ 
and the potential $V(\phi)$ are arbitrary.
However, by adjusting the parameter 
``$\beta$'' and choosing appropriate functionals 
$F(\phi)$ and $V(\phi)$, we can match
the predictions of our inflationary model
with the observations.

The variation of the action \eqref{16} with  
respect to the general metric $g_{\mu\nu}$
defines the Einstein's equation
\begin{equation}
\label{17} 
R_{\mu \nu}-\frac{1}{2}g_{\mu \nu}R =
\kappa T_{\mu \nu}^{({\rm eff})},
\end{equation}
where $T_{\mu \nu}^{({\rm eff})}$ represents
the effective energy-momentum tensor, associated with the
total Lagrangian   
$\mathcal{L}^{({\rm eff})}_{m}
=\beta F(\phi)T+\mathcal{L}_{m}$,
\begin{eqnarray}
\label{2.3}
T_{\mu \nu}^{({\rm eff})} &=&\frac{-2}
{\sqrt{-g}}\frac{\partial(\sqrt{-g}
\mathcal{L}^{({\rm eff})}_{m})}{\partial g^{\mu\nu}}
=T_{\mu \nu}-2\beta F\left(T_{\mu \nu}
-\frac{1}{2}Tg_{\mu \nu}+\Theta_{\mu \nu}\right),
\end{eqnarray}
\begin{equation}
\Theta_{\mu \nu}\equiv g^{\alpha \gamma} 
\frac{\delta T_{\alpha
\gamma}}{\delta g^{\mu \nu}}=-2T_{\mu \nu}+g_{\mu \nu} 
\mathcal{L}_{m}-2g^{\alpha \gamma}\frac{\delta^2
\mathcal{L}_{m}}{\delta
g^{\mu \nu}\delta g^{\alpha \gamma}}.
\end{equation}
The explicit form of the symmetric tensor 
$\Theta_{\mu \nu}$, for the 
real scalar field, possesses the feature 
\begin{equation}
\Theta_{\mu \nu}=-\partial_{\mu}
\phi\partial_{\nu}\phi-T_{\mu\nu}
=-2\partial_{\mu}\phi\partial_{\nu}\phi-g_{\mu\nu}
\left(\frac{1}{2}\dot \phi^2 -V(\phi)\right),
\end{equation}
in which we used a homogeneous inflaton field $\phi=\phi(t)$.
In this case, the Lagrangian \eqref{18} reduces to   
\begin{equation}
\mathcal{L}_{m}=\frac{\dot\phi^2}{2}-V(\phi).
\end{equation}
Note that we have utilized the metric
signature $(-,+,+,+)$. The variation of 
the action \eqref{16} with respect to $\phi$ 
gives the equation of motion of this field
\begin{equation}
\label{2.9}
(1+2\beta F)\nabla_{\mu}\nabla^{\mu}\phi
-(1+4\beta F)\frac{\partial V(\phi)}{\partial\phi}
-2\beta\frac{\partial F(\phi)}{\partial\phi}
\left(2V -\partial_\mu \phi \partial^\mu \phi\right)=0.
\end{equation}

Now we employ a perfect
fluid, represented by $T_{\mu \nu}^{({\rm eff})}=
{\rm diag}\big(-\rho^{{\rm (eff})},
p^{{\rm (eff})},p^{{\rm (eff})},
p^{{\rm (eff})})$, where $\rho^{{\rm (eff})}$ and
$p^{{\rm (eff})}$ show the effective energy
density and effective pressure. 
Thus, these quantities obtain the following forms 
\begin{eqnarray}
\label{21} 
T_{00}^{({\rm eff})}=\rho^{({\rm eff})}=\frac{1}{2}\dot \phi^2 
(1+2\beta F)+(1+4\beta F)V,
\nonumber\\
T_{ij}^{({\rm eff})}=p^{({\rm eff})}g_{ij} =\Big[\frac{1}{2} \dot
\phi^2(1+2\beta F)-(1+4\beta F) V\Big]g_{ij}.
\end{eqnarray}
To determine the explicit form of the equation 
of motion of the scalar field, we can use Eq. \eqref{2.9} and or
the continuity equation $\dot\rho^{({\rm eff})}
+3H\big(\rho^{({\rm eff})}
+p^{({\rm eff})}\big)=0$, which leads
to the generalized Klein-Gordon equation
\begin{equation}
(1+2\beta F)(\ddot \phi+3H\dot\phi)+\beta
F_{,\phi}\dot\phi^2+(1+4\beta F)
V_{,\phi} +4\beta F_{,\phi}V=0.
\end{equation}
where $V_{,\phi}={\rm d} V/{\rm d}\phi$ and $F_{,\phi}
={\rm d} F/{\rm d}\phi$.

By substituting Eqs. \eqref{21} and also the FRW
metric into the Einstein equation \eqref{17}, the Friedmann
equations can be expressed in terms of $\rho^{({\rm eff})}$
and $p^{({\rm eff})}$,
\begin{equation}
\label{23} 
H^2=\frac{\kappa \rho^{({\rm eff})}}{3}= \frac{\kappa}{3}
\bigg[\frac{\dot \phi^2}{2}(1+2\beta F) 
+(1+4\beta F)V\bigg],
\end{equation}
\begin{equation}
\label{24}
\frac{\ddot{a}}{a}=-\frac{\kappa}{6}\big(3p^{({\rm eff})}
+\rho^{({\rm eff})}\big)=-\frac{\kappa}{3}
\bigg[\dot \phi^2(1+2\beta F)-(1+4\beta F)V\bigg],
\end{equation}
\begin{equation}
\label{25} 
{\dot H}=\frac{\ddot{a}}{a}-H^2= -\frac{\kappa}{2}
\dot\phi^2\big(1+2\beta F\big).
\end{equation}
Combining Eq. \eqref{25} with the time
derivative of Eq. \eqref{23} we acquire another 
form of the modified Klein-Gordon equation
\begin{equation}
\ddot \phi+3H\dot\phi(1+2\beta F)+\beta
F_{,\phi}\dot\phi^2+(1+4\beta F)V_{,\phi} +4\beta F_{,\phi}V=0.
\end{equation}

We should note that one may directly extract 
Eq. \eqref{24}, by substituting the FRW metric 
into the action \eqref{16}. Afterward, the variation of the 
resultant action with respect to the degree of 
freedom ``$a(t)$'' gives the equation of motion of $a(t)$.
However, we applied the Einstein's equation 
which is equation of motion for the general 
metric $g_{\mu\nu}$.

The slow-roll approach induces the conditions 
${\dot \phi^2}\ll V$, $\vert\ddot \phi \vert \ll \vert 
3H \dot \phi\vert$
and $F_{,\phi}\dot\phi^2\ll H \dot \phi$. 
Thus, in these approximations, the Friedman
and the modified Klein-Gordon equations
take the features  
\begin{equation}
\label{2.16}
H^2\simeq\frac{\kappa}{3}(1+4\beta F)V,
\end{equation}
\begin{equation}
\label{2.17}
3H\dot\phi(1+2\beta F)+(1+4\beta F)V_{,\phi} +4\beta
F_{,\phi}V\simeq 0.
\end{equation}
In this approximation, the slow-roll parameters
find the forms 
\begin{equation}
\label{30}
\epsilon_V\simeq\frac{1}{2\kappa \big(1+2\beta F\big)}
\left(\frac{V_{,\phi}}{V}+\frac {4\beta F_{,\phi}}{1+4\beta
F}\right)^2,
\end{equation}
\begin{equation}
\label{31}
\eta_V\simeq\frac{1}
{\kappa\big(1+2\beta F\big)}
\left[\frac{V_{,\phi\phi}}{V}
+\frac{2\beta\big(3+4\beta F\big)
F_{,\phi}}{\big(1+2\beta F\big) \big(1+4\beta F\big)}
\frac{V_{,\phi}}{V}+\frac{4\beta
\big(1+2\beta F\big) F_{,\phi
\phi}-8\beta^2F_{,\phi}^2}
{\big(1+2\beta F\big) \big(1+4\beta
F\big)}\right].
\end{equation}
As expected, in the limit $\beta \to 0$,
these parameters prominently reduce to
${\epsilon_{\rm E}}$ and $\eta_{\rm E}$.
 
Another crucial factor for quantifying the inflation
is the number of e-folds $N$. It 
is calculated via the equation 
$\Delta N=N-N_{\rm end}=\int^t_{t_{\rm end}}
H {\rm d}t=\int^{\phi}_{\phi_{\rm end}} 
(H/\dot \phi){\rm d}\phi$. In which 
$N$ represents the number of e-folds at the cosmological
time ``$t$'' during the inflation, 
while $N_{\rm end }$ exhibits
the number of e-folds at the end of the inflation.
In the slow-roll approximation $\Delta N$ is given by 
\begin{equation}
\label{2.20}
\Delta N=\ln\big[a(t_{\rm end})/a(t)\big]
\simeq\int_{\phi_{\rm end}}
^{\phi} \left[\frac{\kappa V}{V_{,\phi}} 
+ \frac{2\kappa\beta V\big(FV_{,\phi}-2F_{,\phi} V\big)}
{V^2_{,\phi}}\right]
{\rm d}\phi,
\end{equation}

The scalar power spectrum $A_s$ is
associated with the curvature perturbations \cite{55}. 
In the slow-roll approximation we receive  
\begin{equation}
\label{2.21}
A_s=\frac{3\kappa H^2}{24\pi^2 \epsilon_V}\simeq
\frac{\kappa^2(1+4\beta F)V}{24\pi^2 \epsilon_V}.
\end{equation}
The Planck, BICEP/Keck, and other observations provide
the following constraint on the amplitude $A_s$ \cite{31},
\begin{equation}
A_s=(2.10\pm0.03)\times 10^{-9}.
\end{equation}

The scalar spectral index is given by 
$n_s=1+\frac{d\ln A_s}{d\ln k}$.
Using Eq. \eqref{2.21} we calculate ``$n_s$'' in terms of 
the slow-roll parameters $\epsilon_V$ and $\eta_V$,
\begin{equation}
n_s \simeq 1+ \frac{d\ln A_s}{d N}=
1+\frac{d\ln A_s}{H dt}=1+\frac{\dot\phi}{H A_s}
\frac{dA_s}{d\phi},
\end{equation}
where $d \ln k\simeq dN$ \cite{6}. Therefore, we obtain 
\begin{equation}
\label{2.24}
n_s=1+\frac{\dot\phi}{H}\Big[\frac{4\beta F_{,\phi}}
{1+4\beta F}+\frac{V_{,\phi}}{V}
-\frac{(\epsilon_V)_{,\phi}}
{\epsilon_V}\Big]=1-2\Big[\epsilon_V
-\frac{(\epsilon_V)_{,\phi}}
{\frac{4\beta F_{,\phi}}
{1+4\beta F}+\frac{V_{,\phi}}{V}}\Big],
\end{equation}
in which we employed Eqs. \eqref{2.16}, \eqref{2.17}
and the definition of $\epsilon_V$ in Eq. \eqref{30}.
By taking into account Eq. \eqref{31}, we receive
\begin{equation}
\label{2.25}
\frac{(\epsilon_V)_{,\phi}}
{\frac{4\beta F_{,\phi}}
{1+4\beta F}+\frac{V_{,\phi}}{V}}=\eta_V-2\epsilon_V
\end{equation}
Thus, ``$n_s$'' takes the following feature  
\begin{equation}
\label{2.26}
n_s \simeq 1+2\eta_V-6\epsilon_V.
\end{equation}

According to the production of the tensor perturbations
in the inflationary period, 
the amplitude of the gravitational waves obtains 
the specific form \cite{49},
\begin{equation}
\label{2.27}
A_t=\frac{2\kappa}{\pi^2}H^2\simeq
\frac{2\kappa^2(1+4\beta F)V}{3\pi^2}.
\end{equation}
From Eq. \eqref{2.27} and \eqref{2.21}, the relationship
between the tensor-to-scalar ratio $r = A_t/A_s$ and
the slow-roll parameter $\epsilon_V$ is given by 
\begin{equation}
\label{2.28}
r\simeq16\epsilon_V.
\end{equation}
The Planck 2018 results \cite{8} provide strong 
limits on the inflationary observables. However, 
the BICEP/Keck 
2021 (BK18) greatly enhance the maximum 
value for the tensor-to-scalar ratio $r$.
The data of the Planck satellite have 
restricted the tensor-to-scalar ratio``$r$'' at 
a $95\%$ confidence level 
\begin{equation}
r<0.065 .
\end{equation}
The BICEP/Keck 2021 (BK18) \cite{56} impose the 
following restriction
\begin{equation}
r<0.036 .
\end{equation}
The preferred range for the scalar spectral index ``$n_s$'' 
via the Planck+BK18 is
\begin{equation}
n_s=0.964\pm0.004.
\end{equation}

The tensor spectral index ``$n_t$'',
based on its definition $\frac{d\ln A_t}{d k}$, 
can be rewritten as
\begin{equation}
\label{2.29}
n_t\simeq\frac{d\ln A_t}{d N}=\frac{\dot\phi}{H A_t}
\frac{dA_t}{d\phi}=-\frac{1}{\kappa(1+2\beta F)}
\left[\frac{4\beta F_{,\phi}V+(1+4\beta F)
V_{,\phi}}{(1+4\beta F)V}\right]^2=-2\epsilon_V.
\end{equation}
Thus, we acquire the relation $n_t=-r/8$. 
This is consistent with the 
results in the framework of the general relativity for a
single field under the slow-roll approximation
\cite{6}, \cite{49}. Notably, since the
ratio ``$r$'' always is positive, ``$n_t$'' 
possesses only negative values.

We should note that 
the slow-roll approximation is typically utilized in the
computation of the inflationary properties.
Using the slow-roll approximation
for the general relativity models 
implies that the slow-roll
parameters are modest with respect to unity. Some of 
these models include the scalar fields that 
are minimally coupled.

\section{The D-brane inflation}
\label{400}

The D-brane inflation models have 
two main aspects: a string-theoretical appearance
and a phenomenological view. Both 
are very interesting, specially after 
investigating the inflationary models in the Planck 2018 data
\cite{31}. The Planck 2018 $n_s-r$ plane is shown
in the figure 1. 

\begin{center}
\begin{figure}
\centering
\includegraphics[width=14cm]{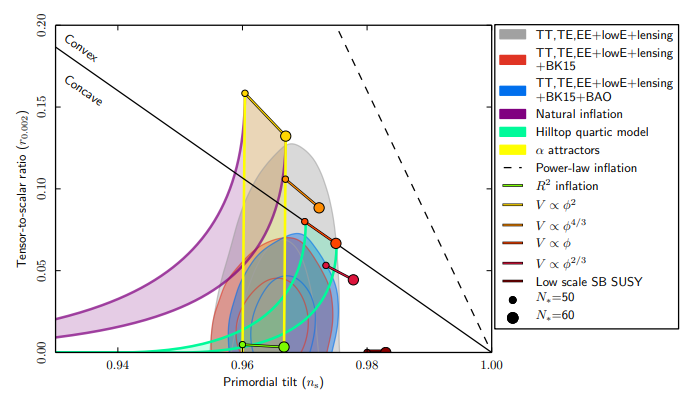}
\caption{\scriptsize{Using the Planck data
alone (the grey area), or with the BICEP2/Keck data 2014 (red),
and BAO (blue) data. The marginalized joint 68\% and
95\% CL areas for ``$n_{s}$''
and ``$r$'' at $k=0.002 Mpc^{-1}$ were
compared with the theoretical
predictions of specific inflationary theories. It should be
noted that ${\rm d}n_{s}/{\rm d}\ln k = 0$ is assumed 
in the combined 68\% and 95\% CL areas. The lines
display the predictions of several models as a function
of the number of e-folds $N_{*}$, 
until the end of the inflation.}}
\end{figure}
\end{center}

The dark blue (light blue) region describes the
1$\sigma$ (2$\sigma$) confidence level for the CMB
data, obtained by the Planck 2018 and BICEP/Keck 2014.
They also include the baryon oscillations (BAO) data.
Meanwhile, the comparison of the predictions of inflationary
models with the data in \cite{31} is on the basis of the CMB
data only, excluding the BAO. The corresponding
1$\sigma$ and 2$\sigma$ regions are shown in  
red. As observed in \cite{31}, the left-hand side of the
1$\sigma$ dark blue and also the red regions  
are covered by the two yellow lines,
corresponding to the $\alpha$-attractors. 
Its predictions are corresponding to
$n_s=1-2/N$, which have been shown via two yellow lines. 
This investigation is usually focused
on the configurations which contain the e-fold 
numbers $N$ = 50 and $N$ = 60. The $\alpha$-attractor models 
include various inflationary models
\cite{3}, \cite{38}, \cite{57}, \cite{58}.
 
As an example, let us simultaneously plot the predictions
of the $\alpha$-attractors and of 
the simplest D-brane inflationary
model with $V\thicksim1-\big(\frac{m}{\phi}\big)^4$.
The figure 2 represents the Planck 2018 data for ``$r$'' on
$\rm log_{10} r$ scale. This is more suitable for
illustration of the predictions of the models in the
limit of small ``$r$''. Both of these classes
of models exhibit an attractor behavior. As one
can see from this figure, the combination of the simplest
$\alpha$-attractor model and the simplest D-brane inflation
model almost completely covers the $1\sigma$ dark blue
(dark purple) area of the Planck 2018 data.

\begin{center}
\begin{figure}
\centering
\includegraphics[width=10cm]{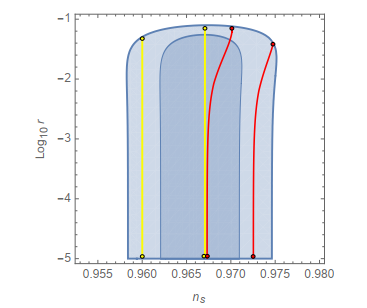}
\caption{\scriptsize{The Planck 2018 results \cite{8} for 
``$n_s$'' and ``$r$'' have been compared with 
the predictions of the simplest D-brane inflationary
model with $V\thicksim1-(\frac{m}{\phi})^4$ 
and the $\alpha$-attractors in the $2\sigma$ region. 
The dark (light) blue region on the panel
represents the Planck 2018 $1\sigma$ ($2\sigma$) region.
The data are related to the CMB. On the panel,
the quadratic $T$-model of the $\alpha$-attractors at 
$N = 50$ and $N = 60$ has been represented 
by the two yellow lines. Besides,
the simplest D-brane inflation model has been 
represented by the two red lines.}}
\end{figure}
\end{center}

The D-brane inflation model after the 
Planck 2018 has phenomenologically
acquired a new significance. The string 
theory origin of the D-brane inflation 
model is often attributed to the model in \cite{21}, 
in which the ($\overline{\rm D3}$-brane)-(D3-brane)
interaction has been studied. 
Predictions of this model completely cover the 
right-hand side of the 1$\sigma$ dark blue 
and also dark red areas in figure 1.
The inflationary potentials, 
corresponding to the ($\overline{\rm D3}$-brane)-(D3-brane)
interaction, have been proposed as in the following forms 
\begin{equation}
\label{3.3}
V_{\rm BI}=V_0\left[1-\left(\frac{m}{\phi}\right)^{7-p}+
\cdots \right],
\end{equation}
\begin{equation}
\label{3.4}
V_{\rm KKLTI}=V_0\left[1+\left(\frac{m}{\phi}
\right)^{7-p}\right]^{-1},
\end{equation}
where BI (KKLTI) represents the Brane Inflation
(KKLT Inflation). In \cite{16} a cosmological 
analysis of these potentials has been 
performed, with emphasis on the D3- and 
D5-branes. A consistent cosmological evolution
of the D3-branes in the string theory, with 
the 10-dimensional geometry, can be interpreted as an 
evolution in the 4-dimensional spacetime under the condition 
that the 6-dimensional internal space 
possesses a constant volume. 

In accordance with the above explanation
the D3-brane model can be defined via the following 
potential 
\begin{equation}
\label{3.5}
V(\phi) = V_0\left(1-\frac{m^4}{\phi^4}\right).
\end{equation}
Here we eliminated the higher-order terms which  
represent additional stabilizing
terms. They are unimportant during the inflation.
In other words, they become significant only after
the inflaton. 

Our suggested functional $F(\phi)$ is as follows
\begin{equation}
F(\phi)={\frac {{\phi}^{5}}{\mu} \left
( 1-{\frac {{m}^{4}}{{\phi}^{4}}} \right)}.
\end{equation}
This form of $F(\phi)$ incorporates
a non-minimal interaction between the gravity and matter.
In fact, we have extended the functional
$F(\phi)$, which was initially expressed 
as $F(\phi) \sim \phi$ in \cite{48}. 
However, for $\phi^2\ll m^2$ the functional
$F(\phi)$ is approximately proportional to $\phi$. 
As $\phi$ tends to zero, this choice of $F(\phi)$
obviously leads to the Einstein's theory of gravity,
i.e. $F(0)=0$. 

More detail for the theoretical justification of the 
forms of $V (\phi)$ and $F (\phi)$, via the 
physical and string-theoretic motivations,
is given by the following descriptions.

\textbf{1}. \textbf{Justification for the potential form}

This potential can be motivated from the several 
perspectives:

\textbullet \ In the D-brane inflation models, the 
inflaton $\phi$ often represents the separation 
in a higher-dimensional space between a D-brane 
and an anti-D-brane (or another D-brane). They 
have an attractive potential that resembles 
Coulomb's form \cite{21}:
\begin{equation}
V(\phi) = V_0\left(1-\frac{C}{\phi^n}\right),
\end{equation}
where $C$ is a constant, associated with the D-brane 
tension and couplings, and ``$n$'' is dependent on 
the dimensions of the interacting D-branes and the 
compactified space. The leading-order 
interaction for the D3-branes in the 10-dimensional 
Type IIB string theory is proportional to ${1}/{\phi^4}$, 
which supports our selection of $n=4$.

\textbullet \ The Moduli Stabilization Effects: 
The $m^4$-term may be extracted from the non-perturbative 
effects that stabilize the moduli fields, such as the gaugino 
condensation or the instanton corrections. These  
introduce a little correction to the flat potential 
at large $\phi$ \cite{59}.

\textbullet \ Comparison with the Plateau Potentials: 
Such potentials are favored by the Planck data, because 
they produce sufficient inflation while keeping 
``$r$'' small, which is consistent with the observations.

\textbf{2}. \textbf{Justification for the form of $F(\phi)$ }

The functional $F(\phi)$ that we have 
used, generalizes the linear case 
$F(\phi)=\phi$. It includes the higher-
order of $\phi$, introduced by the  
higher-order modifications to the effective 
string actions. This form of 
$F(\phi)$ preserves the same symmetry and 
dimensional structure of the potential. It was 
designed to reflect the shape of the tachyon 
potential, which is compatible with the brane 
collapse scenarios.

\textbullet \ The stringy Non-minimal Coupling: In 
the string theory, non-minimal couplings between 
the scalars (like the dilaton or moduli) and the matter 
(encoded in $T$) are common. The $\phi^5$-term 
could arise from the higher-derivative corrections 
(e.g., $\alpha^{\prime}$ corrections in the string 
effective actions) \cite{60}.

\textbullet \ The factor $(1-\frac{m^4}{\phi^4})$ 
ensures the regular behavior near a specific cut-off 
scale (the brane separation or resolution radius), 
which is commonly used to mimic the stringy corrections 
or the moduli stabilization effects in the UV.

\textbullet \ The UV-Completion and cut-off Scale 
$\mu$: The parameter $\mu$ acts as a cut-off 
scale, suppressing the higher-dimensional operators. 
This is natural in the effective field theories, 
derived from the string compactifications. Thus, 
$\mu$ could be related to the string scale $M_s$ 
or the Kaluza-Klein scale.

\textbullet \ Matching to the Linear Limit: For 
$\phi\gg m$, we obtain $F (\phi ) \approx {\phi^5}/{\mu}$, 
but if we tune $\mu\sim \phi^4$ (or take into account 
a regime in which $\phi$ is small), it simplifies to 
$F(\phi)\sim\phi$. Hence, it reduces to the simpler 
form \cite{48}. This ensures consistency 
with the earlier studies.

\textbf{3}. \textbf{Theoretical Consistency 
and the Effective Field Theory Validity }

\textbullet \ The model ensures that the corrections 
are perturbative by assuming $\phi\gg m$ during 
inflation. 

\textbullet \ Since $T$ (the trace of the stress-energy 
tensor) is a scalar under the diffeomorphisms, and the 
coupling respects the general covariance, the coupling 
term $F(\phi)T$ is consistent with the effective field 
theory principles.

Consequently, the D-brane inflation in the string 
theory, where the inflaton reflects the branes separation 
and the potential results from the brane-antibrane 
interactions (mostly $\propto 1/\phi^4$ for the
D3-branes), serves as the motivation for the forms 
of $V(\phi)$ and $F(\phi)$. Using $\phi^5/\mu$ to 
capture the higher-order stringy corrections and $\mu$ 
as a UV cut-off, the coupling term $F(\phi)T$ generalizes 
minimal couplings, looked at in the previous work. The 
stabilization effects of the moduli are encoded by the 
$m^4/\phi^4$ term. These choices guarantee the Planck 
data compatibility while extending the simpler 
models \cite{61}. 

\textbf{4}. \textbf{ Motivation for the Specific 
Choice of $f(\phi, T)$}

The scalar field $\phi$ is coupled to the trace of the 
energy-momentum tensor $T$ in the $f(\phi,T)$ 
framework, which generalizes the scalar-tensor theories. 
This coupling term is inspired by the following criteria.

\textbullet \ Dynamics of the D-branes: In the string theory, 
the metric and the scalar field (such as the inflaton) 
are coupled to the D-branes. 
The D-brane action in a warped throat geometry 
naturally gives rise to the expression $F(\phi)T$.

\textbullet \ Modified gravity: $F(\phi)T$ includes 
a direct interaction between the inflaton and the  
matter/radiation sectors, which can change the 
dynamics of the inflation and reheating, in contrast to 
the pure $f(R)$ or the Brans-Dicke theories.

\textbullet \ Connection to the Conformal Invariance:

The specific form $F(\phi) = \frac {{\phi}^{5}}{\mu} 
\left ( 1-{\frac {{m}^{4}}{{\phi}^{4}}} \right) $ 
has been chosen 
because it reduces to $F(\phi)\sim\phi$ (a common 
conformally invariant coupling) in the limit 
$\phi \gg m$, which links it to the well-studied models.

\textbullet \ The correction term ($1-\frac{m^4}{\phi^4}$) 
and the deviation from linearity, i.e. ($\phi^5$), are driven by 
the D-brane potentials in the flux compactifications (e.g., KKLT 
scenarios) \cite{62}, where the inflaton potential is 
corrected by the non-perturbative effects or the supersymmetry 
breaking.

\textbf{5}. \textbf{Connection Between the D-brane Inflation 
and $f(\phi,T)$ Theory}

In the D-brane inflation, the inflaton $\phi$ represents the 
position of a D-brane in a warped throat (e.g., the 
Klebanov-Strassler throat) \cite{63}. The brane's 
motion is governed by a scalar potential $V(\phi)$ 
(e.g., from the anti-D-brane tension or the flux-induced terms). 
Couplings to the bulk energy-momentum tensor $T$, 
arise from the brane's interaction with the bulk 
fields (e.g., via the DBI action) \cite{20}.
This interaction is encoded by the $f(\phi,T)$-term; 
the brane's movement causes the perturbations to the 
matter fields and the bulk geometry. This results in a 
$T$-dependent correction to the 4-dimensional effective action. 
Our $F(\phi)$ is obtained from 
the warp-factor dependency and the non-canonical 
kinetic terms of the D-brane action.

By substituting the functionals \eqref{3.5} 
and  (3.4) into
Eqs. \eqref{30} and \eqref{31}, the slow-roll 
parameters find the features
\begin{eqnarray}
\epsilon_V&\simeq& - \,{\frac {8\mu\, \left
( 3\,\beta\,{m}^{8}\phi+2\,
\beta\,{m}^{4}{\phi}^{5}-5\,\beta\,
{\phi}^{9}-{m}^{4}\mu \right) ^{2}
}{\kappa \phi^2 \left( {m}^{4}-{\phi}
^{4} \right) ^{2} \left( 4\,\beta\,{
m}^{4}\phi-4\,\beta\,{\phi}^{5}
-\mu \right) ^{2}\, \left( 2\,
\beta\,{m}^{4}\phi-2\,\beta\,
{\phi}^{5}-\mu \right) }},
\end{eqnarray}
\begin{eqnarray}
\eta_V&\simeq&-\frac{1}{ \kappa 
\phi^2 \left( {m}^{4}-
{\phi}^{4} \right) \left( 4\,\beta\,
{m}^{4}\phi-4\,\beta\,{\phi}^{5}-\mu
 \right)  \left( 2\,\beta\,{m}^{4}
\phi-2\,\beta\,{\phi}^{5}-\mu
 \right) ^{2}}
\nonumber\\
&\times& 4\mu\Bigg( 30 {\beta}^{2}
{m}^{12}{\phi}^{2}-
50 {\beta}^{2}{m}^{8}{\phi}^{6}+10 
{\beta}^{2}{m}^{4}
{\phi}^{10}+10 {\beta}^{2}{\phi}^{
14}-24 \beta {m}^{8}\mu \phi
\nonumber\\
&+&20 \beta {m}^{4}\mu {\phi}^{5}-20 
\beta \mu {\phi}^{9}+5 {m}^{4}
{\mu}^{2} \Bigg).
\end{eqnarray}

The quantities ``$r$'' and ``$n_s$'' can be expressed 
in terms of the parameters ``$\mu$'', ``$\beta$'' and ``$m$''
and the inflaton field $\phi$, 
\begin{eqnarray}
r&=& \frac{8}{\kappa\Bigg[1+\frac{2 \beta
{\phi}^{5} \left
(1-{\frac {{m}^{4}}{{\phi}^{4}}} \right) 
}{\mu}\Bigg]}{\Bigg(\frac{4\,{m}^{4}
}{{\phi}^{5} \left( 1-{\frac {{m}^{4}}
{{\phi}^{4}}} \right)}+\frac{\frac{4\beta}{\mu}
\bigg[ 5\,{\phi}^{4} \left( 1-{\frac {{m}^{4}}
{{\phi}^{4}}}
\right) +4\,{m}^{4}\bigg]
}{ 1+\frac{4 \beta \phi^5 \big( 1-\frac
{m^4}{\phi^4}\big)}{\mu}
}\Bigg)^2},
\end{eqnarray}
\begin{eqnarray}
n_{s}&=&1+\frac{8}{\kappa\Bigg[1+\frac
{2 \beta {\phi}^{5}
\left( 1-{\frac {{m}^{4}}{{\phi}^{4}}} \right) 
}{\mu}\Bigg]}
\nonumber\\
&\times& \Bigg\{-\frac{5 {m}^{4}
}{{\phi}^{6} \left( 1-{\frac {{m}^{4}}{{\phi}^{4}}} \right) 
}+\frac{\frac{2 \beta  {m}^{4}}{\mu} \Big[ 3+4 {\frac 
{\beta {\phi}^{5}}{\mu} \left( 1-{
\frac {{m}^{4}}{{\phi}^{4}}} \right) } 
\Big] \Big[5\phi^4\big(1-\frac{m^4}{\phi^4}\big)+4m^4\Big]
}{\left( 1-{\frac {{m}^{4}}{{\phi}^{4}}} \right) 
\Big[1+4 {\frac
{\beta {\phi}^{5}}{\mu} \left( 1-{\frac {{m}^{4}
}{{\phi}^{4}}} \right) } \Big] 
\Big[ 1+2 {\frac {\beta {\phi}^{5
}}{\mu} \left( 1-{\frac {{m}^{4}}{{\phi}^{4}}} 
\right) } \Big] {\phi}^{5} }
\nonumber\\
&+& \frac{ \frac{20\beta}{\mu}\Big[ 1+2 {\frac 
{\beta {\phi}^{5}}{\mu} \left( 1-{
\frac {{m}^{4}}{{\phi}^{4}}} \right) } 
\Big]\Big[\phi^3 (1-\frac{m^4}{\phi^4})+{
\frac {{m}^{4}}{\phi}} \Big]
-2 \frac{\beta^2}{\mu^2} 
\Big[5\phi^4 \big(1-\frac{m^4}{\phi^4}\big)+4m^4 \Big] ^{2}
}{ \Big[1+4 {\frac {\beta {\phi}^{5}}{\mu} 
\left( 1-{\frac {{m}^{4}
}{{\phi}^{4}}} \right) } \Big]  
\Big[ 1+2 {\frac {\beta {\phi}^{5
}}{\mu} \left( 1-{\frac {{m}^{4}}{{\phi}^{4}}} 
\right)}\Big]}
\nonumber\\
&-&6\Bigg[\frac{m^4
}{{\phi}^{5} \left( 1-{\frac {{m}^{4}}{{\phi}^{4}}} 
\right) }+\frac{\frac{\beta}{\mu}\Big[ 5\phi^4
\big(1-\frac{m^4}{\phi^4}\big)+4m^4 \Big]
}{1+4 {\frac {\beta {\phi}^{5}}{\mu} 
\left( 1-{\frac {{m}^{4}}{{\phi}^
{4}}} \right) }}\Bigg]^2\Bigg\}.
\end{eqnarray}
The $\phi$ represents the field value at the time
of horizon crossing, which is determined numerically from
Eq. \eqref{2.20}. The value $\phi_{\rm end}$ is specified 
by either $\epsilon_V=1$ or $\eta_V=1$. 
In this case, $N$ is fixed at 50 or 60 e-folds.
By adjusting appropriate values for the parameters, 
it is feasible to receive any desirable values 
for ``$n_s$'' and ``$r$''.

We have employed a modified gravity 
model that not only aligns with 
the predictions on the right-hand side
of the Planck data but also extends 
to encompass the left-hand side too.
In the case of $\beta=-0.008$ and 
$m=0.1$, the value of ``$\mu$'' should
tend towards higher numbers. Its 
range may begin at $\mu=30$ and can 
extend to $\mu=200$ or higher. Within 
this range, there are $50\leq N \leq 60$. In this case, we
can cover the left-hand side of the 
Planck data. With $\beta=-0.001$
and $m=0.1$, we receive the values of ``$r$'' and ``$n_s$''
in the ranges $r=\left[23\times10^{-6},
39\times10^{-6}\right]$ and $n_s=[0.9517,0.9639]$ 
in which $5 \leq \mu \leq 30$.
Thus, the left-hand side of the Planck data for $N=50$
is effectively covered. 

Now consider the table 1.
 
\newpage
\begin{table}[h]
\caption{The tensor-to-scalar ratio ``$r$'' and the scalar 
spectral index ``$n_s$'' have been numerically calculated  
for $N=50 \to 60$, with $m=0.08$, $\beta=-0.004$ 
and varying $\mu$.} 
\centering 
\begin{tabular}{c rrrr} 
\hline\hline 
$N$&$\mu$& $n_s\quad\quad\quad$& $r\quad
\quad\quad$&  \\ [0.5ex]
\hline 
50$ \rightarrow$60 & 15& 0.9537$ 
\rightarrow$0.9583 & 0.000019$ 
\rightarrow$0.000011\\ 
50$ \rightarrow$60 & 20& 0.9567$ 
\rightarrow$0.9615 & 0.000022$ \rightarrow$0.000014\\
50$ \rightarrow$60 & 25& 0.9586$ 
\rightarrow$0.9635 & 0.000024$ \rightarrow$0.000016\\
50$ \rightarrow$60 & 30& 0.9599$ 
\rightarrow$0.9649 & 0.000025$ \rightarrow$0.000017\\ 
50$ \rightarrow$60 & 35& 0.9608$ 
\rightarrow$0.9659 & 0.000026$ \rightarrow$0.000018\\
50$ \rightarrow$60 & 40& 0.9615$ 
\rightarrow$0.9666 & 0.000027$ \rightarrow$0.000018\\
50$ \rightarrow$60 & 45& 0.9621$ 
\rightarrow$0.9672 & 0.000027$ \rightarrow$0.000019\\
50$ \rightarrow$60 & 50& 0.9625$ 
\rightarrow$0.9677 & 0.000028$ \rightarrow$0.000019\\
[1ex] 
\hline 
\end{tabular}
\label{tab:hresult}
\end{table}

Beside this table, choosing $1\leq\mu \leq 5\;$, 
$\beta=-0.004$ and $m=0.02$ allows us to effectively cover 
the left-hand side of the Planck data. In this
case, for $N=50$ the values of ``$r$'' and ``$n_s$'' 
belong to the intervals $\left[3\times10^{-6} , 
4\times10^{-6} \right]$ and $\left[0.9544 , 0.9640\right]$,
respectively. In the same way, for $N=60$ there are
$r \in [10^{-6}, 3\times10^{-6}]$ and 
$n_s \in [0.9591,0.9693]$.

Now look at the table 2. 
\begin{table}[h]
\caption{The tensor-to-scalar ratio ``$r$'' and the scalar 
spectral index ``$n_s$'' have been numerically calculated 
for $N=50 \to 60$, with $m=0.3$, $\beta=-0.0001$ 
and varying $\mu$.} 
\centering 
\begin{tabular}{c rrrr} 
\hline\hline 
$N$&$\mu$& $n_s\quad\quad\quad$& $r\quad
\quad\quad$&  \\ [0.5ex]
\hline 
50$ \rightarrow$60 & 7 & 0.9567$ 
\rightarrow$0.9615 & 0.000128$ 
\rightarrow$0.000082\\
50$ \rightarrow$60 & 9 & 0.9588$ 
\rightarrow$0.9637 & 0.000140$ \rightarrow$0.000093\\ 
50$ \rightarrow$60 & 11 & 0.9602$ 
\rightarrow$0.9652 & 0.000148$ \rightarrow$0.000101\\
50$ \rightarrow$60 & 13 & 0.9611$ 
\rightarrow$0.9662 & 0.000154$ \rightarrow$0.000106\\
50$ \rightarrow$60 & 15 & 0.9618$ 
\rightarrow$0.9670 & 0.000158$ \rightarrow$0.000107\\
50$ \rightarrow$60 & 17 & 0.9624$ 
\rightarrow$0.9676 & 0.000162$ \rightarrow$0.000113\\
50$ \rightarrow$60 & 19 & 0.9628$ 
\rightarrow$0.9680 & 0.000164$ \rightarrow$0.000116\\
50$ \rightarrow$60 & 21 & 0.9632$ 
\rightarrow$0.9684 & 0.000166$ \rightarrow$0.000118\\
[1ex] 
\hline 
\end{tabular}
\label{tab:hresult}
\end{table}

In addition to the above table, 
the choices $3 \leq \mu \leq 100$,
$\beta = -0.001$ and $m = 0.08$ cover the left-hand side of
the Planck data. Consequently, for $N=50$, we obtain 
$r\in \left[16 \times 10^{-6} , 31 \times 10^{-6}\right]$ 
and $n_s\in \left[ 0.9507 ,0.9661\right]$.
For the case $N=60$, we receive 
$r\in \left[9\times10^{-6}, 23\times10^{-6}\right]$ and 
$n_s \in \left[0.9552,0.9716\right]$. 
The D-brane inflation with a single-field 
model in the Einstein's gravity incorporates 
the red region of the Planck/BICEP/Keck data.
Appropriate values of the parameters
``$\beta$'', ``$\mu$'' and ``$m$'' enabled us 
to encompass the region in the $(n_s, r)$-space 
that was preferred by the Planck 2018. 

Now consider the figure 3.

\newpage
\begin{center}
\begin{figure}
\centering
\includegraphics[width=16cm]{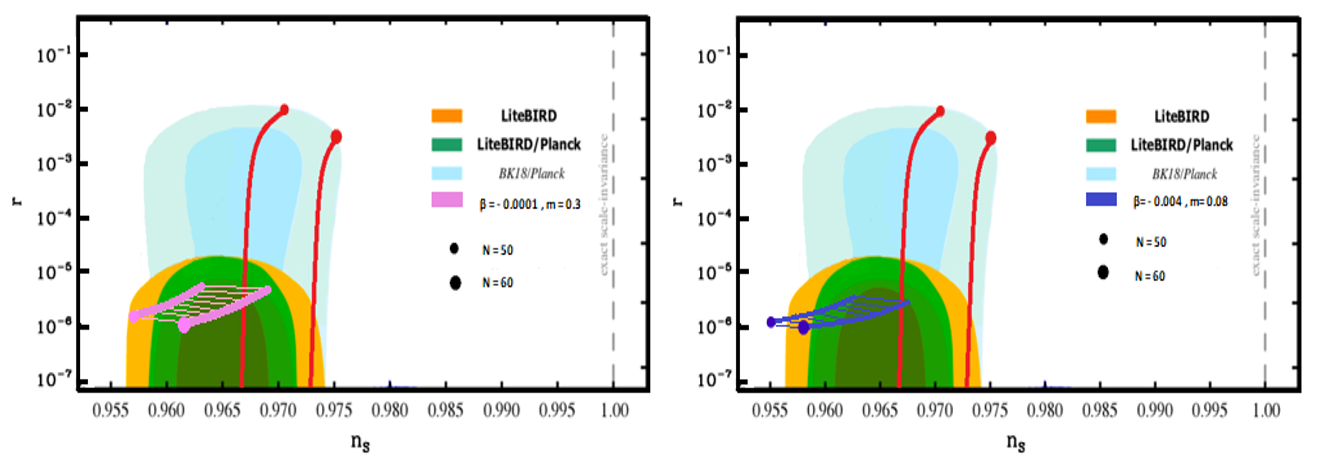}
\caption{\scriptsize{The predicted ``$r$'' and ``$n_s$''
in our model. To compare, the 
outcome of the Einstein's gravity (indicated by the red bar) 
has been displayed, which corresponds
to the D-brane inflationary model 
with $V\sim1-\frac{m^4}{\phi^4}$.
The blue color is assigned when $\beta = -0.004$, $m = 0.08$
and $15 \leq \mu \leq 50$, while 
the pink color is assigned when
$\beta = -0.0001$, $m = 0.3$ and $7 \leq \mu \leq 21$. 
At $k=0.002 Mpc^{-1}$, ``$r$'' and ``$n_s$'' 
are characterized by the confidence level
margins, depicted in the dark (light) blue, corresponding to
the 1$\sigma$ (2$\sigma$) regions, respectively. The 2018
Planck data \cite{8} encompasses the related data 
to the Einstein's gravity.}}
\end{figure}
\end{center}

According to this figure, which is inferred from the 
calculations, we can conclude that by choosing 
appropriate values for the parameters in the D-brane
inflation, we can cover the left-hand side of the Planck data
in the $(n_s ,r)$-plane. Other suitable data enable us
to cover the right-hand side of the data, like the 
Einstein's gravity. Finally, we have reexamined our 
model predictions for the tensor-to-scalar ratio in 
the light of the recent observational updates from the 
BICEP/Keck 2021 experiment. We find that the 
values that we obtained for $r$ and $n_s$ are consistent 
not only with the Planck 2018 data, which include 
these constraints, but also with the $r< 0.036$ 
$(95\% CL)$ bound in the BICEP/Keck 2021 
(BK18) data. This indicates that our model is 
consistent with the observations for a wide range 
of the initial conditions and the parameter choices.

\textbf{$\bullet$} \textbf{ Reheating 
Dynamics in the Modified Gravity}

Next, we shall examine the reheating and its
corresponding dynamics. The oscillation 
of the inflaton field $\phi$ 
around its minimum occurs after the end of the inflation. 
The decay rate $\Gamma_{\phi}$ indicates 
that the energy is transferred into the radiation through 
the decay. The equations for the evolution of the radiation 
energy density and the inflaton energy density are as follows
\begin{eqnarray}
&\dot\rho_{\phi}&+\ 3H(1+\omega_{\phi})\rho_
{\phi}=-\Gamma_{\phi}\rho_{\phi}, \nonumber\\
& & \dot\rho_r+4H\rho_r=\Gamma_{\phi}\rho_{\phi},
\end{eqnarray}
where $\omega_{\phi}$ is the effective equation 
of state during the oscillations, often it vanishes for 
the quadratic-like oscillations, which gives 
\begin{equation}
H_{\rm reh} \sim \Gamma_{\phi}.
\end{equation}

Due to the modifications, at reheating we have 
\begin{equation}
\rho_{\rm total}=\frac{3}{\kappa}\ \frac{H^2}{1+
4\beta F(\phi_{\rm reh})}.
\end{equation}
The universe thermalizes with a radiation bath 
at the moment when H $\sim \Gamma_{\phi}$,
\begin{equation}
\rho_r=\frac{\pi^2}{30}g_{*}T^4_{\rm reh},
\end{equation}
and
\begin{equation}
\rho_r\approx\frac{3}{\kappa}\ \frac{\Gamma_
{\phi}^2}{1+4\beta F(\phi_{\rm reh})}.
\end{equation}
Thus, the temperature at the end of reheating is 
defined as follows
\begin{equation}
T_{\rm reh}=\bigg(\frac{90}{\pi^2g_{*}\kappa}\ \frac
{\Gamma_{\phi}^2}{1+4\beta F(\phi_{\rm reh})}\bigg)^{1/4},
\end{equation}
where $g_{*}$ is an effective relativistic degree of 
freedom, and its typical value is $g_{*}\sim$100. Near 
the minimum of the potential ($\phi\sim m$), the 
reheating temperature simplifies as
\begin{equation}
T_{\rm reh}\approx \bigg(\frac{90}{\pi^2g_{*}\kappa}
\Gamma_{\phi}^2\bigg)^{1/4}.
\end{equation}

Assuming a constant $\omega_{\phi}$ and a thermodynamic 
transition. Therefore, the number of e-folds during 
the reheating is defined as follows
\begin{equation}
N_{\rm reh}=\frac{1}{3(1+\omega_{\phi})} \ln \bigg
(\frac{\rho_{end}}{\rho_{\rm reh}}\bigg),
\end{equation}
where $\rho_{\rm end}\sim V(\phi_{\rm end})$ and $\rho_
{\rm reh}=\frac{\pi^2}{30}g_{*}T^4_{\rm reh}$. 
Hence, we receive  
\begin{equation}
N_{\rm reh}=\frac{1}{3(1+\omega_{\phi})} \ln \bigg[\frac
{V(\phi_{end})}{\frac{\pi^2}{30}g_{*}T^4_{\rm reh}}\bigg],
\end{equation}
\begin{equation}
\omega_{\rm reh}=\frac{1}{3N_{\rm reh}} \ln \bigg[\frac
{V(\phi_{end})}{\frac{\pi^2}{30}g_{*}T^4_{\rm reh}}\bigg]-1.
\end{equation}

The figures 4, 5, 6 and 7 are an analysis of these equations. 
For each figure, we have shown separately 
the range of our parameters. The reheating 
must occur after the end of the inflation, so the equation 
of state should satisfy $-1/3<\omega<1$. The plots 
cover $10^5-10^{15}$ GeV, safely above the the lower bound
of the Big-Bang Nucleosynthesis ($\gtrsim 1$ MeV). 
The range of 1-60 e-folds 
is plausible, though the realistic scenarios typically favor 
$N_{\rm reh}\lesssim 20$. The plots correctly show shorter 
reheating for stiffer $\omega_{\rm reh}$ at 
the fixed $T_{\rm reh}$. 
Therefore, the plots are physically reasonable if 
$\omega_{\rm reh}>-1/3$ is enforced. The reheating 
temperatures and durations align with the theoretical 
expectations, and no immediate conflicts arise 
from the presented ranges. These results help 
to quantify viable reheating scenarios within 
the observational and thermodynamic constraints.

\begin{center}
\begin{figure}
\centering
\includegraphics[width=16cm]{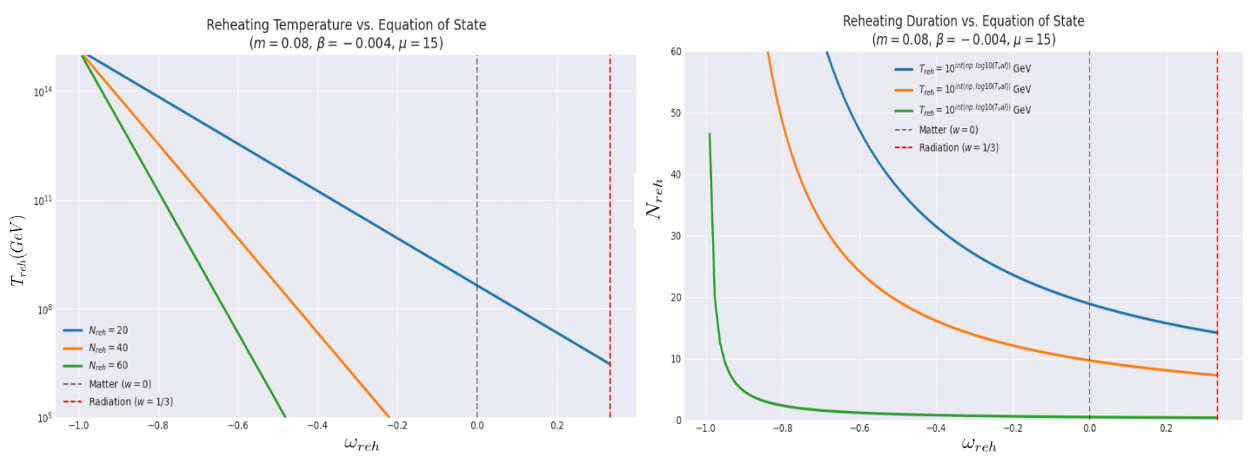}
\caption{\scriptsize{The left plot illustrates 
that how the universe's temperature at the end 
of the reheating depends on the cosmic fluid's 
behavior, parameterized by $\omega_{\rm reh}$, 
which determines the expansion rate. For a 
fixed number of the e-folds $N_{\rm reh}$, a faster 
expansion (higher $\omega_{\rm reh}$) dilutes 
the energy and induces a low temperature. 
The dashed lines mark the key cases: $\omega_{\rm reh}=0$ 
(matter-like) and $\omega_{\rm reh}=-1/3$ 
(radiation-like). The right plot shows that how 
$N_{\rm reh}$ varies with $\omega_{\rm reh}$ for the fixed 
reheating temperatures. A softer equation of 
state ($\omega_{\rm reh}\to -1/3$) slows the energy 
dilution, which permits a longer reheating. A 
stiffer state requires the shorter reheating to 
achieve the same temperature. This reveals an 
interplay between the expansion dynamics and the  
thermal history post-inflation.}}
\end{figure}
\end{center}
\begin{center}
\begin{figure}
\centering
\includegraphics[width=11cm]{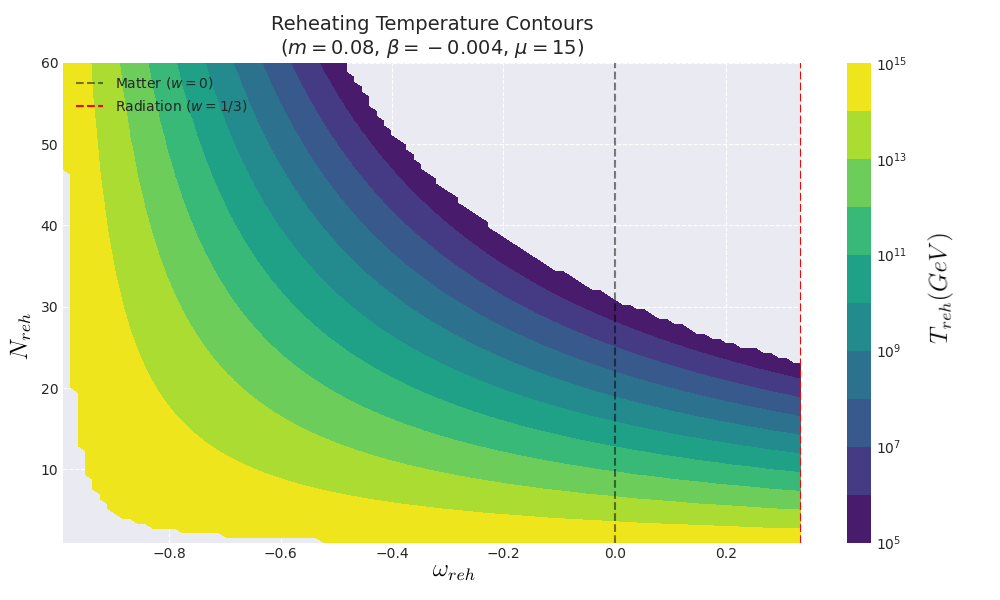}
\caption{\scriptsize{This contour plot provides a 
comprehensive visualization of the reheating 
temperature landscape, shaped by the 
effects of the reheating equation 
of state and the duration. Given the energy 
density at the end of the inflation, the temperature 
after the reheating sensitively depends on how 
the Universe expands and cools during this 
epoch. The regions with the higher $\omega_{\rm reh}$ 
correspond to the faster dilution of the energy density, 
which yield the lower reheating temperatures for the 
same reheating length. The interplay manifests 
as nonlinear contours and reflects the exponential 
scaling of the energy density with $ N_{\rm reh}$ and 
equation of state. This reveals the physically plausible 
reheating scenarios, consistent with the theoretical 
inflation models.}}
\end{figure}
\end{center}
\begin{center}
\begin{figure}
\centering
\includegraphics[width=16cm]{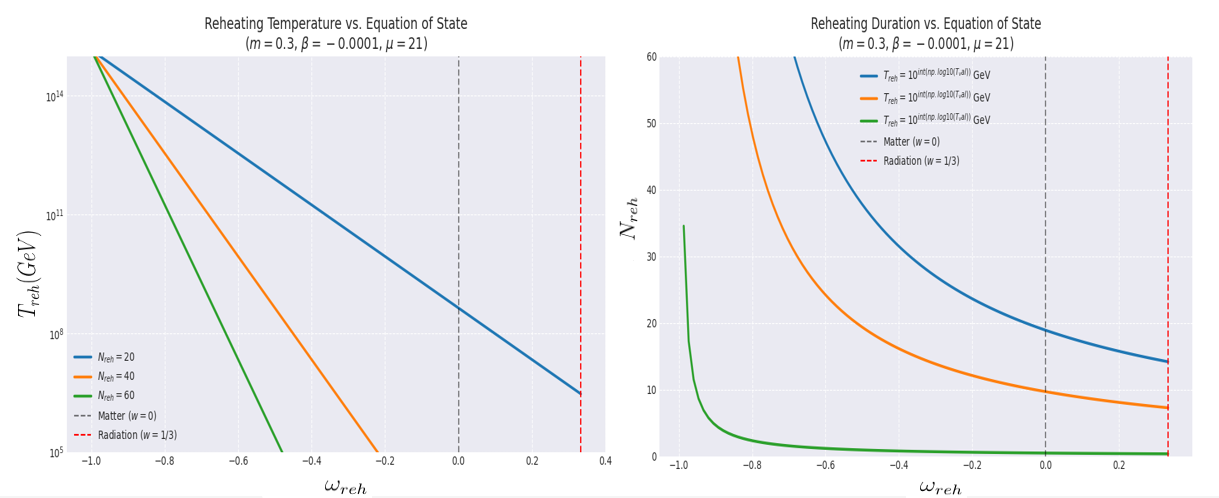}
\caption{\scriptsize{The left figure shows that how the 
reheating temperature varies with the equation of 
state parameter $\omega_{\rm reh}$ . The different  
durations ($N_{\rm reh}$) are 
fixed. A larger $\omega_{\rm reh}$ 
(with the range from $-1$ to $1/3$) means the faster energy 
dilution, which reduces the temperature. The key case-matter 
($\omega_{\rm reh}=0$) and the radiation domination 
($\omega_{\rm reh}=1/3$) have been highlighted. 
They illustrate that 
how the reheating dynamics shapes the early universe's 
thermal state. The right figure demonstrates that, 
for a given reheating temperature, a stiffer equation 
of state (higher $\omega_{\rm reh}$) requires fewer 
e-folds ($N_{\rm reh}$) due to rapid energy loss, while 
the softer equations of state need the longer reheating. 
The boundaries at $\omega_{\rm reh}=0$ and $1/3$ 
mark the standard reheating scenarios, i.e. linking the 
expansion behavior to the post-inflation thermal history.}}
\end{figure}
\end{center}
\begin{center}
\begin{figure}
\centering
\includegraphics[width=11cm]{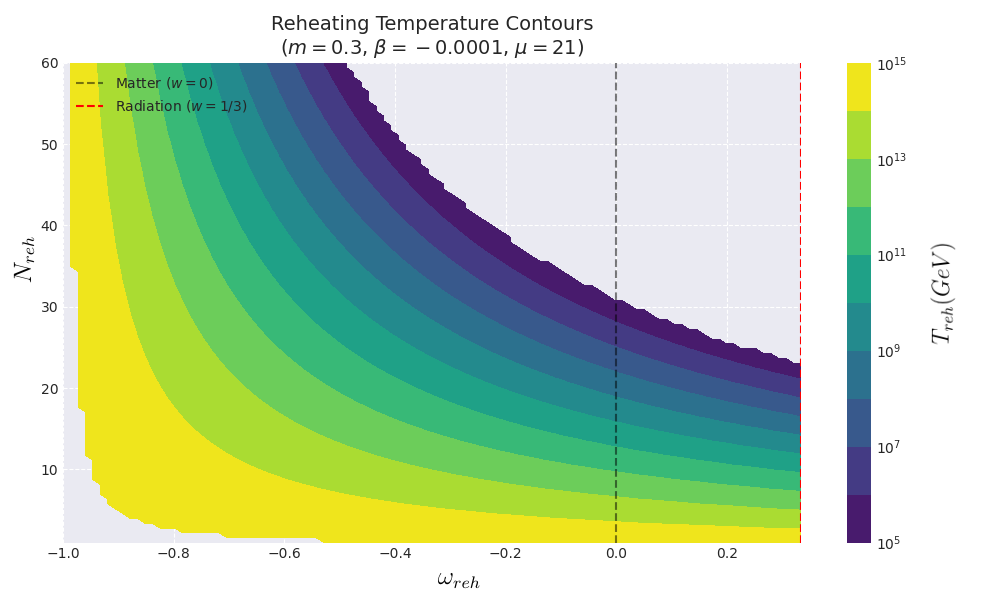}
\caption{\scriptsize{This contour map consolidates the 
dependence of the reheating temperature on both the 
equation of state and the reheating 
duration. This illustrates that 
the full parameter space has been 
spanned by $\omega_{\rm reh}$ 
and $N_{\rm reh}$. The color gradation (log scale) 
demonstrates how the interplay between the expansion 
dynamics (controlled by $\omega_{\rm reh}$) 
and the reheating 
time determines the final thermal state of the Universe. 
The regions of the high reheating temperature correspond to 
the relatively short reheating epochs with the soft equations 
of state, whereas the cooler reheating temperatures 
arise for the longer reheating durations or the stiffer 
equations of state. This plot provides a comprehensive 
overview of the reheating phase, which is consistent with 
the specified inflationary energy scale and the model 
parameters.}}
\end{figure}
\end{center}

The figures 4-7 present an analysis of two parameter sets, 
with the similar trends, expected 
for other values (not shown). 
Due to the wide range, selected 
for $T_{\rm reh}$, the figures 
appear nearly identical. The parameters $\beta$, $m$ 
and $\mu$ demonstrate comparable effects on the 
reheating dynamics in both cases. While the calculated 
reheating parameters ($T_{\rm reh}$, $N_{\rm reh}$ and 
$\omega_{\rm reh}$) show the measurable differences 
between the sets. Because of the broad scales using,
these variations are visually subtle in the plots. 
Nevertheless, all graphical results remain fully 
consistent with the underlying reheating physics.

\textbf{$\bullet$} \textbf{ Discussion of the Limitations}

In the context of the large-field inflationary models, the 
slow-roll condition fails when the inflaton field value 
is $\phi \lesssim m$. This is corresponding to the 
inflationary phase's termination. This behavior aligns 
precisely with the theoretical constraints, derived from 
brane-antibrane separation dynamics \cite{62}. At 
substantially larger field values ($\phi \gg m$), the 
scalar potential enjoys protection through the shift 
symmetry, a characteristic property of the axion-like 
fields in the string-theoretic framework \cite{60}.

The higher-order perturbative corrections and 
non-perturbative effects may gain relevance when 
$\phi \sim m$. However, their contributions are naturally 
regulated by the intrinsic cut-off scale $\mu$, 
embedded in the functional form of $F( \phi)$ \cite{22}. 
The specific coupling structure $f(\phi,T)=\beta F(\phi)T$ 
necessarily mediates direct energy transfer between the 
oscillating inflaton field $\phi$ and the fundamental 
matter fields during the reheating epoch.

The efficiency of the reheating mechanism exhibits 
dual dependence on: (i) the dimensionless coupling 
parameter $\beta$, and (ii) the complete spectrum of the  
available inflaton decay channels. Although the  
potential gradient near $\phi \sim m$ suggests an 
efficient energy transfer, a comprehensive 
treatment requires an explicit specification of the matter 
sector Lagrangian $L_m$. It should include particular 
couplings to the Standard Model constituents such 
as the Higgs field \cite{64}.

\section{Conclusions}
\label{500}

We included the new term $F(\phi)\;T$
in the Einstein-Hilbert action, 
motivated by the $f(R, T)$ gravity.
We initially focused 
on the inflationary dynamics through the 
arbitrary functionals $F(\phi)$ and $V(\phi)$.
Afterward, we computed the 
cosmological observables ``$r$'' and ``$n_{s}$'' for
the special forms of $F(\phi)$ and $V(\phi)$.
Therefore, we produced appropriate data corresponding
to the D-brane inflation. Then, we 
compared the accuracy and precision
of our data with the experimental data. 

We applied the following numerical values for the
coupling constant, i.e., $\beta = - 0.004$ 
and $- 0.0001$, and the parameter ``$\mu$'' was left as 
a variable to obtain ``$r$'' and
``$n_s$''. All the data have been shown in the tables 1 and 2,
and the paragraphs immediately after these tables.
Our method covered the left-hand side of the 
Planck's data, as it was depicted in the figures 2 and 3.
By choosing suitable 
numerical values for ``$\beta$'' and ``$\mu$''
we can also cover the entire Planck data surface.

The figure 3 displays the results for two situations.
In the first case, which was depicted 
by the blue color, the parameters have been selected as 
$m = 0.08$, $\beta = -0.004$
and $15 \leq \mu \leq 50$. As a
result, the data remain in the 
designated region of the Planck
data in the $(n_s, r)$-space plot as 
blue. In the second case,
the parameters for the pink part are 
$m =0.3$, $\beta = -0.0001$ and $7 \leq \mu \leq 21$. 
The pink points in the $(n_s, r)$-space stay in the
$2\sigma$ region. By choosing some specific
values for the parameter ``$\mu$'' from the range 
$3 \leq \mu \leq 100$, we can cover 
the left-hand side of the Planck data.
Typically, if we choose a wider 
range for the parameters ``$\beta$'' and ``$\mu$'', we can 
completely cover all the sweet spots of the Planck data. 

We compared our results with the standard simplest D-brane
inflation in the Einstein's gravity. As 
it is clear in the figure 3, our results
completely agree with the Planck data. 
The case $\beta = 0$ elaborates the identical 
results with the standard simplest D-brane 
inflation in the Einstein's gravity.

We comprehensively analyzed the 
dynamics of the reheating phase via the 
four figures 4 - 7. 
The presented results in these figures are fully consistent 
with both our theoretical interpretation of the reheating 
and the underlying equations that describe this epoch. 
The graphical representations directly reflected the 
physical behavior, derived from our computations.
They are completely consistent with the coherent correspondence 
between the analytical predictions and the numerical results.


\end{document}